\documentclass[11pt,a4]{article}

\usepackage[dvips]{graphics}
\usepackage{epsfig}

\newcounter{ctr}

\begin{document}

\title{
Origin of complexity in multicellular organisms}
\author{ Chikara Furusawa and
        Kunihiko Kaneko \\
        {\small \sl Department of Pure and Applied Sciences}\\
        {\small \sl University of Tokyo, Komaba, Meguro-ku, Tokyo 153, JAPAN}
\\}
\date{}
\maketitle

\begin{abstract}
Through extensive studies of dynamical system 
modeling cellular growth and reproduction, 
we find evidence that complexity arises in 
multicellular organisms naturally through evolution.
Without any elaborate control mechanism, these systems can exhibit 
complex pattern formation with spontaneous cell differentiation.
Such systems employ a `cooperative' use of
resources and maintain a larger growth speed
than simple cell systems, which exist in a homogeneous state 
and behave 'selfishly'.
The relevance of the diversity of chemicals and reaction dynamics to the growth of
a multicellular organism is demonstrated.
Chaotic biochemical dynamics are found to provide the multi-potency of stem cells.
\end{abstract}

Multicellular organisms consist of differentiated cell types, with rather
complex biochemical dynamics exhibited by complex metabolic and genetic networks.  
Through the course of development, cells differentiate into several types and often form 
complex patterns. The molecular mechanisms existing of each stage of development have
been elucidated experimentally \cite{Cell}. 
However, it is still not clear why multicellular organisms should have such complexity, 
nor why such inhomogeneities in cell types and patterns are common.
These are  not trivial problems \cite{Smith}, since
a simple biochemical network would be sufficient (and perhaps more fit) to produce identical cells rapidly and faithfully.
Here we give an answer to the problem of why multicellular organisms in 
general have diverse cell types with complex patterns and dynamics.

Indeed, by considering dynamical systems of appropriate biochemical networks, 
it is possible to show that inhomogeneous patterns with diverse cell types can 
emerge. Such work was pioneered by Turing \cite{Turing,Newman}, and  extended 
in Refs. \cite{IDT1,IDT2,IDT3}. Differentiation into several cell types has also 
been studied by using random genetic networks \cite{Kauffman}.
Still, it is not known how biochemical and genetic networks allowing for
diversity are evolved.

Any theory accounting for such cellular diversity requires a specific choice
of the genetic and biochemical reaction networks and parameters. Then, to explain 
the ubiquitous nature of cellular diversity in multicellular organisms, 
we must understand how such reaction dynamics allowing for cell 
differentiation can evolve, without postulating a 
finely tuned mechanism. 
Here we study the relationship between the growth of 
multicellular organisms and some characteristics of intra-cellular dynamics, to understand the origin of cellular diversity.
Since the question we address is rather general, we seek an answer that does not rely on
detailed knowledge of specific cellular processes in existing organisms.
Instead, we adopt dynamical systems models that display only essential 
features of the developmental process.  In a minimal model of
cellular dynamics, we include only intra-cellular biochemical
reactions with enzymes, simple cell-cell interactions through diffusion,
and cell division as a result of biochemical reactions within each cell.  
By studying a class
of models with this general and minimal content, we show that 
cell differentiation from a stem cell and development producing a
complex pattern are natural consequences of evolution.

{\bf Model:}
First, cells are assumed to be completely surrounded by a one-dimensional medium.
The state of a cell $i$  is assumed to be characterized by the cell volume and 
$c^{(m)} _i(t)$, the concentrations of $k$ chemicals($m=1,\cdots,k$).  
The concentrations of chemicals change as a result of
internal biochemical reaction dynamics within each cell and
cell-cell interactions communicated through the surrounding medium. 
The corresponding chemical concentration in the medium is denoted by 
$C ^{(m)} (x,t)$, where $x$ denotes the position 
along the one-dimensional axis.
We use a one-dimensional model only for its tractability; 
conclusions we draw in this case are consistent with the result of 
preliminary simulations of a two-dimensional model.

{\bf Internal reaction dynamics:} 
For the internal chemical reaction dynamics, we choose a catalytic network among 
the $k$ chemicals, represented by a matrix $Con(\ell,j,m)$ 
which takes on unity when the reaction from the chemical $\ell$ to $m$ is catalyzed by $j$, and 0 otherwise.
The rate of increase of $c^{(m)}_i(t)$ (and decrease of $c^{(\ell)}_i(t)$) 
through this reaction is given by $c^{(\ell)}_i(t) (c^{(j)}_i(t))^{\alpha}$, where
$\alpha$ is the degree of catalyzation($\alpha=2$ in most simulations here).
Each chemical has several paths to other chemicals, and thus a complex 
reaction network is formed.

Besides, we take into account
the change in the volume of a cell, that varies as a result of transportation of
chemicals into the cell from the environment.  
For simplicity, we assume that the volume is proportional to the sum of 
chemicals in the cell.  
The concentrations of chemicals are diluted as a result of an increase in the
volume of the cell.  With the above assumption, this dilution effect
is tantamount to imposing the restriction $\sum_{\ell} c^{(\ell)}_i=const.$  

Of course, real biochemical processes within a cell are much more 
complicated.  Instead of taking such details into account,
we employ this simple model, and consider a variety of reaction 
networks created randomly, to elucidate a general 
relationship between the growth of cells and cellular dynamics.

{\bf Cell-cell interactions through a medium:}
Cells interact with each other through the transport of chemicals out of and into 
the surrounding medium.  As a minimal case, we consider only indirect 
cell-cell interactions through diffusion of chemicals via the medium.
The diffusion coefficient should depend on the
chemical.  Here, we assume that there are two types of chemicals, one
which can penetrate the membrane and one which can not.  We use the
notation $\sigma_m$, which takes the value 1 if the chemical
$c^{(m)}_i$ is penetrable, and 0 otherwise.

To sum up all these process, the dynamics of chemical concentrations in each 
cell is represented as follows:

\begin{equation}
dc^{(\ell)}_i(t)/dt  =  \Delta c^{(\ell)}_i(t) - c^{(\ell)}_i(t) \sum_{l=1}^k\Delta c^{(\ell)}_i(t) \;,
\end{equation}
with 
\begin{eqnarray}
&\Delta c^{\ell}_i(t)=\sum_{m,j}Con(m,j,{\ell}) \;e_1 \;c^{(m)}_i(t) \;(c^{(j)}_i(t))^\alpha \nonumber & \\
& \mbox{~~~~ } - \sum_{m',j'} Con({\ell},j',m') \;e_1 \;c^{({\ell})}_i(t) \;(c^{(j')}_i(t))^\alpha  \nonumber & \\
& \!\!\! + \sigma_{\ell} D(C^{(\ell)} (p^x_i,t) -
c^{({\ell})}_i (t))\;. &
\end{eqnarray}

The variable $p^x_i$ denotes the 
location of the $i$-th cell.  
The second term in Eq.~(1) represents the dilution effect by changing the volume of the cell.
On the other hand, the diffusion of chemicals in the medium are governed by 
the following equation for $C^{(\ell)}$:

\begin{eqnarray}
&\partial C^{(\ell)} (x,t)/\partial t=-\tilde{D}\nabla ^2 C^{(\ell)}
(x,t) \nonumber & \\
& +\sum_i\delta (x-p^x_i)\sigma_{\ell} D(C^{(\ell)} (x,t)
-c^{({\ell})}_i (t))\;
\end{eqnarray}

where the boundary condition is chosen to be $C(0,t)=C(x_{max},t)=const.$, 
$\tilde{D}$ is the diffusion constant of the environment, and $x_{max}$ 
denotes the extent of the medium.
This boundary condition can be interpreted 
as a chemical bath outside of the medium, which supplies
those penetrable chemicals that are consumed to the medium via a
constant flow to the cell.

{\bf Cell Division:}

Each cell takes penetrable chemicals from the medium as the nutrient,
while the reaction in the cell transforms them to unpenetrable
chemicals which construct the body of the cell.
As a result of these reactions, the volume of the cell is
increased.
In this model, the cell is assumed to divide into two almost
identical cells when the volume of the cell is doubled.

During this division process, all chemicals are almost equally divided,
with tiny random fluctuations (e.g., $\sim 10^{-6}$ $c^{(\ell)}_i$).  
The case of equal division is assumed, since
we do not intend to introduce an elaborate mechanism, while the fluctuations
are introduced because we wish to study development processes that 
are robust with respect to molecular fluctuations. 
After cell division, two daughter cells appear around 
their mother cell's position, 
and the positions of all cells are adjusted so that the distances between 
adjacent cells are 1.
As a result, the total length of the chain of cells increases.

As the initial state, a single cell, whose chemical concentrations are determined randomly, is placed.
According to the process described above, cells divide to form a chain.  
Of course, the behavior of the model 
depends on each intra-cellular reaction network.  We have carried out 
simulations of the model by considering 800 different reaction networks, generated randomly.
Throughout the paper, the number of chemical species $k$ is 20, and each chemical
has 6 reaction paths to other chemicals, chosen randomly.
Each chemical reaction path is catalized by other (or the same) chemical, 
again chosen randomly.  Among the 20 chemicals,
3 chemicals can penetrate cell membranes.
The parameter are fixed at $e_1=1.0, D=1.0, \tilde{D}=2.0$, and $x_{max}=500$.

As results, we found that each growth curve using different reaction network can be classified into two classes:
(I) `fast' growth in which the increase of cell number grows exponentially in time $t$, 
and (II) `slow' growth in which the cell number grows linearly in time. 
Under the parameters presented above, approximately 5\% of randomly chosen reaction networks show the case (I) behavior.
These two classes, indeed, are also distinguished by 
the nature of the corresponding cellular dynamics.

In case (II), the chemical compositions and dynamics of all cells 
are identical.  These dynamics fall into either a fixed point
or a limit-cycle attractor.
In this situation, only a few cells around the edges of the chain can divide.
Since cells are not differentiated, chemicals required for cell 
growth are identical for all cells.  Thus, once the cells at the edges consume the required
resources, the remaining cells can no longer grow.  This is the reason for the
linear growth.

In case (I) with exponential growth, it is found that
the chemical reaction dynamics of the cells are more complex than in case (II).
The microscopic differences between the chemicals in 
two daughter cells are amplified through the internal biochemical 
dynamics and the cell-cell interaction, leading to chaotic chemical dynamics.
For most such cases, the cells differentiate into various 
cell types that are defined by distinct chemical dynamics 
and compositions (see Fig.1).
Even though no external mechanism of differentiation is implemented,
the instability that results from the intra-cellular dynamics and 
the cell-cell interactions brings about transitions to various cell types \cite{max_num}.
Indeed, this differentiation is a general feature of a system of interacting units that each possesses some non-linear internal dynamics \cite{GCM1,GCM2}, 
as has been clarified by isologous diversification theory \cite{IDT1,IDT2,IDT3}.

In case (I), the division of cells is not restricted to the edge of the chain.
There is a flow of the nutritive chemicals into the inside of the chain, that can maintain the growth of internal cells. 
This flow is sustained by the diffusion between cells
possessing different chemical compositions and exhibiting different phases
of chemical oscillations.
Here, the capability for cell division is also differentiated as the development 
progresses. The total increase in cell number is due to the division of 
certain type(s) of cells, and most other cell types stop dividing.

In the developmental processes in case (I), 
initial cell types exhibit transitions to other types.
Cells of the initial cell types either proliferate or switch to other types
stochastically.
They can be regarded as a stem cell \cite{Ogawa}.  
The dynamics of a stem-type cell, that can produce
different cell types after divisions, exhibit
chaotic oscillation(see Fig.1, and also\cite{IDT3}).  
The type of differentiated cells maintain their type after divisions. 

Next, we compare the growth of a single cell with that of an ensemble of cells.
In Fig.2 the growth speeds of a single cell and an ensemble of cells are 
plotted.
Here each point corresponds to a different reaction network. 

The points around the peak of the growth for an ensemble correspond to case (I) of exponential growth. 
Here, the growth speed of a single cell is not large.
In each cell, a variety of chemicals coexist, supporting
complex reaction dynamics and cell differentiation.
In case (I), organisms with cellular diversity, plotted by green points, have a larger growth speed
of an ensemble than those without diversity.  Indeed, in the former case, cells
differentiate in the use of nutrients to take them efficiently.

In case (II), there are much simpler intra-cellular reaction processes, with only a few auto-catalytic reactions used dominantly, that can produce the rapid replication of a single cell.
In this case, the growth speed of a single cell is often large (represented by some of the blue points in Fig.2), 
while the growth speed of an ensemble always remains small.
In some sense, simple cells with rapid growth are `selfish':
Although such simple cells with low diversity of chemical species
can exhibit large growth speeds as single cells, 
they cannot grow cooperatively, and their growth speeds as ensembles are
suppressed because of strong competition for resources.

The developmental process presented in case (I) is irreversible.
Initially, cells have complex chaotic internal dynamics and 
a variety of chemicals.  They have the  potential to differentiate into 
cell types with simpler cellular dynamics, for example, with a 
fixed-point and regular oscillation (see Fig.1 as an example).
This type of cell with simpler dynamics can produce cells of the same type
or stop dividing.  Hence, the loss of multipotency,
known to exist in the developmental process of real organisms \cite{Cell}, 
is explained in terms of the decrease in time of diversity in the 
intra-cellular dynamics and chemicals.
Indeed, the decrease in complexity of dynamics with the loss of multipotency has been 
confirmed quantitatively, for example by computing 
the Kolmogorov-Sinai(KS) entropy of the intra-cellular dynamics.
Using the data in Fig.2, we have found that 
case (I) has a positive KS entropy while for the case (II) it is zero.  
Furthermore, there is positive correlation between the KS entropy and 
the growth speed as an ensemble.
The decrease of KS entropy in going from a stem cell to
other differentiated cells has also been confirmed.
For example, the KS entropy of the stem-type ``red'' cells in Fig.1 is approximately $2.4\times 10^{-4}$, while it is $3.1\times 10^{-5}$ for ``green'', $4.5\times 10^{-5}$ for ``blue'', less than $1.0\times 10^{-5}$ for ``yellow'' cells.

Our results are robust against the change in model parameters, as long as
the internal reaction and intra-cellular diffusion terms have a comparable order of
magnitude.  The results are also independent of the details of the model, in particular,
of the choice of catalyzation degree $\alpha=1$ or 3.
The same two classes and the same relation between the growth and cellular diversity are obtained,
as long as the number of chemicals is sufficient (say for $k>10$)  and number of
the reaction paths is in the medium range (e.g., 3$\sim$9 for each chemical at $k=20$; 
otherwise, the intra-cellular dynamics
fall onto a fixed point for most cases without differentiation).

To sum up, our study has provided evidence that an ensemble of cells with 
a variety of dynamics and stable states (cell types)
has a larger growth speed than an ensemble of simple cells 
with a homogeneous pattern, because of the greater 
capability of the former to transport and differentiation in the use of nutritive chemicals.
Note that no elaborate mechanism is required for the appearance of such heterogeneous cell ensembles.
Some fraction of the randomly chosen biochemical networks we considered exhibit dynamics sufficiently complex to allow for spontaneous cell differentiation.

Our result suggests that complexity of multicellular organisms
with differentiated cell types is a necessary course in evolution, once a multicellular
unit emerges from cell aggregates.  In fact, by carrying out the
evolution experiment numerically, with mutation to reaction networks and selection
of the cell ensembles with higher growth speed,  we have confirmed that
cells of the case (I) emerge and survive through evolution.

Our result concerning the relationship between the diversity of chemicals and
dynamics and the growth speed of a single cell and ensemble
provides experimentally testable predictions.
Since even primitive organisms such as {\sl Anabena} \cite{Anabena1}
and {\sl Volvox} \cite{Volvox} exhibit differentiation in cell types
and some spatial pattern,
the relationship can be verified.
In fact, for a mutant of {\sl Volvox} that possesses only homogeneous cells, 
the growth becomes slower in comparison to the wild-type \cite{Volvox2}.  

Chaotic intra-cellular dynamics and the diversity of chemicals present in 
a stem cell are also experimentally verifiable.  Decrease of diversity in 
chemical composition and of complexity in their temporal variation is
expected with the decrease of multipotency of a cell.

We would like to thank Tetsuya Yomo for stimulating discussions.
This research was supported by
Grants-in-Aid for Scientific
Research from the Ministry of Education, Science, and Culture
of Japan (11CE2006)

\begin{figure}[htbp]
\begin{center}
\includegraphics[width=15cm,height=12cm]{./fig1.eps}
\end{center}
\caption{
An example of the developmental process with exponential growth.
(a) The developmental process of the spatial pattern with differentiated cells,
starting from a single cell.
Up to several divisions, a single type of cell reproduces, maintaining its 
characteristic type of dynamics.
This type is represented as ``red" cells.
The time series of the chemical concentrations of this type of cell are plotted in (b), where the time series of only 6 chemical concentrations among the 20 are overlaid, to avoid indistinct figure. (The other chemicals exhibit similar dynamics for each).
With further increase of cell number, some of the ``red'' cells start to 
exhibit different types of chemical dynamics, due to some influence of 
the cell-cell interaction on the intra-cellular dynamics.
The transitions from the ``red'' cells to three distinct cell types
are represented by the plots of ``green", ``blue", and ``yellow" cell types in (c), (d), and (e).
The time series of concentrations of the 6 chemicals are again plotted.
In this example, only these 4 types
of cells appear in the developmental process, while no intermediate
types exist.  
Here, the ``red'' cells are regarded as the stem cells that have the potential 
both to reproduce themselves and to differentiate into other cell types, while 
the differentiated (``green'', ``blue'', and ``yellow'') cells have lost such potential 
and only reproduce the same type.  
In this example, the growth of the organism is due to the division of ``red'' cells, while ``green'' and ``yellow'' cells keep their volume and ``blue'' cells gradually shrink.
A part of the spatial pattern
formed by these 4 types of cells is shown in (a).
}
\end{figure}

\begin{figure}[htbp]
\begin{center}
\includegraphics[width=13cm,height=13cm]{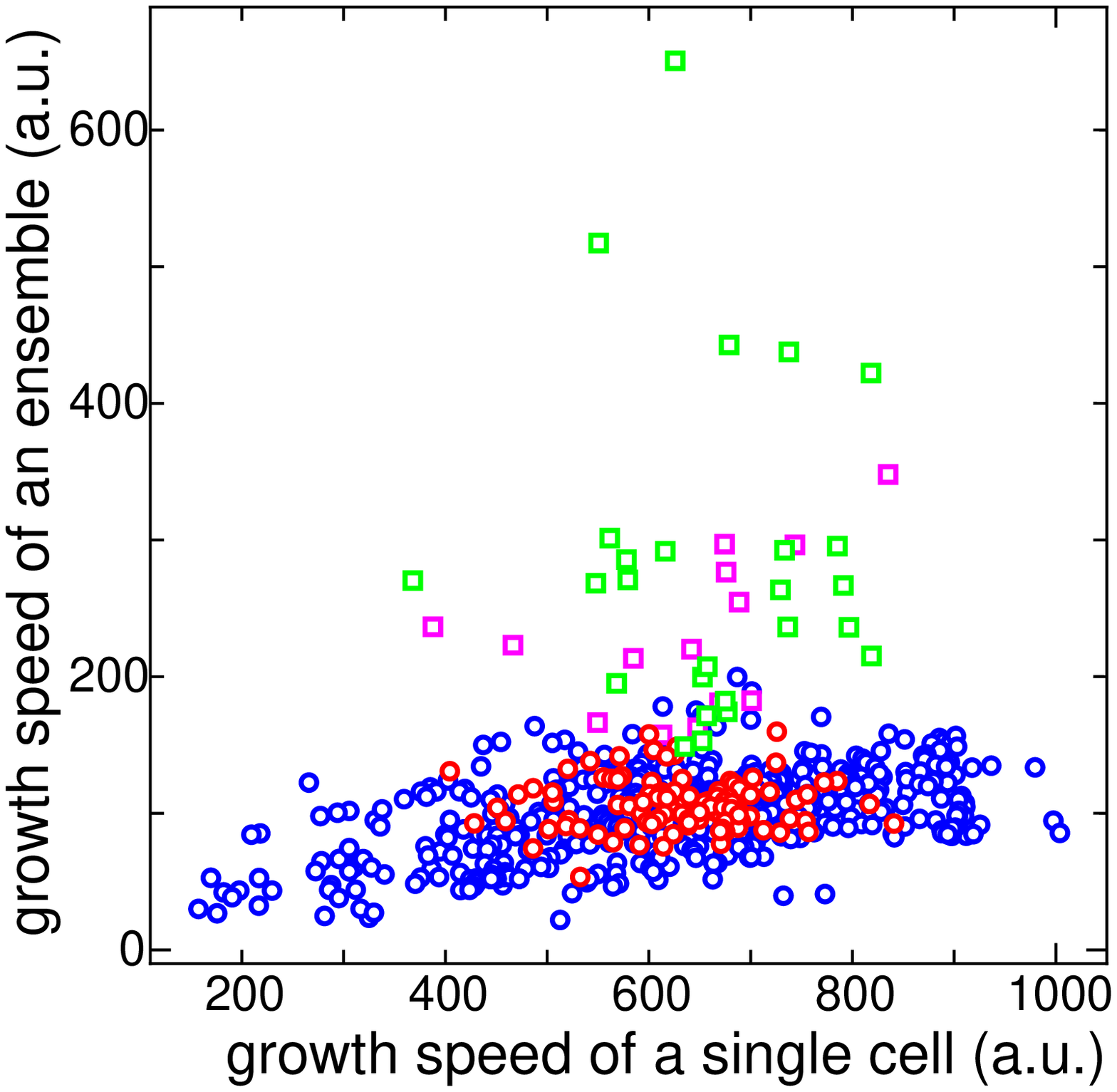}
\end{center}
\caption{
Relationship  between the growth speed of a single cell
and that of an ensemble.  The ordinate shows the growth speed
of an ensemble, measured as the inverse of the
time for the cell number doubles from 100 to 200, while the abscissa represents 
the inverse of the time for a single cell to divide.
Each point is obtained by using a different chemical reaction network.
The blue circles correspond to case (II), where the state of cells are identical 
and the growth of an ensemble is linear in time.  
The dynamics of the chemical concentrations
of these cases are mostly of the fixed point type, with a few cases of 
limit cycles.
The green and red rectangles correspond to the case 
with chaotic cellular dynamics, where the growth of an ensemble is exponential
in time.  The green points correspond to the case with cell differentiation into
several types.  For the red points, cell differentiation into
distinct types is not observed (up to the time when the cell number reaches 200).
}
\end{figure}


\begin{thebibliography}{999}
\bibitem{max_num}
Under the adopted parameters, we have observed up to 6 types of cells.
As the number of chemicals is increased, the number of cell types slowly increase, although it is hard to define the cell types in such high-dimensional cellular dynamics and measure the maximal number of distinct cell types.
\end{thebibliography}
\end{document}